\documentclass[aps,pre,twocolumn]{revtex4}
\usepackage{amsmath,bm,epsfig}

\begin{document}
\title{Detecting and localizing the foci in human epileptic seizures}
\author{Eshel Ben-Jacob$^1$, Stefano Boccaletti$^{2,3}$, Anna
Pomyalov$^3$, Itamar Procaccia$^3$ and Vernon L. Towle$^4$}
\affiliation{$^1$ School of Physics and Astronomy, Tel Aviv University, Tel Aviv 69978, Israel\\ 
$^2$ Embassy of Italy in Tel Aviv, Trade Tower, 25 Hamered St., Tel Aviv 68125, Israel\\ 
$^3$ Department of Chemical Physics, The Weizmann Institute of Science, Rehovot 76100, Israel\\
$^4$ Dept. of Neurology, The University of Chicago, 5841 S. Maryland Ave., Chicago, IL 60637}

\date{\today}
\begin{abstract}
We consider the electrical signals recorded from a subdural
array of electrodes placed on the pial surface of the brain for
chronic evaluation of epileptic patients before surgical
resection. A simple and computationally fast method to analyze the interictal
phase synchrony between such electrodes is introduced and
developed with the aim of detecting and localizing the foci of the
epileptic seizures. We evaluate the method by comparing the results
of surgery to the localization predicted here. We find an
indication of good correspondence between the success or failure in
the surgery and the agreement between our identification and the
regions actually operated on.
\end{abstract}
\maketitle

{\bf Epilepsy is defined as a condition of chronic unprovoked
seizures.  It is one of the most common neurologic disorders, affecting
more than 50 million people worldwide.  About 1/3 of cases are not
adequately controlled by medication, causing a severe reduction in
independence and quality of life.
 If an individual's seizures are stereotyped and appear to have a
focal onset, surgical resection may be a consideration. Localizing
accurately the epileptic focus (or foci) is therefore essential part
of surgical planning.  We propose a simple and computationally fast
method based on phase synchronization, that is able to correctly
localize the foci of epileptic seizures. The method was applied to the
data of three patients. The overlap of the location of the
electrodes identified by our method and the resected areas was
consistent with the degree of surgery success.}

\section{Introduction}

 Epilepsy is defined as a condition of chronic unprovoked
seizures.  It is one of the most common neurologic disorders, affecting
more than 50 million people worldwide.  About 1/3 of cases are not
adequately controlled by medication, causing a severe reduction in
independence and quality of life. Seizures often begin in one brain
region and rapidly spread to other areas of cortex, manifesting in a
wide variety of symptoms, ranging from a brief staring spell (absence
seizures), to generalized convulsions (grand mal seizures). The mechanisms of
seizure spread are not yet well  understood.  If an
individual's seizures are stereotyped and appear to have a focal
onset, surgical resection may be a consideration.

\begin{figure}
\epsfig{width=.35\textwidth,file=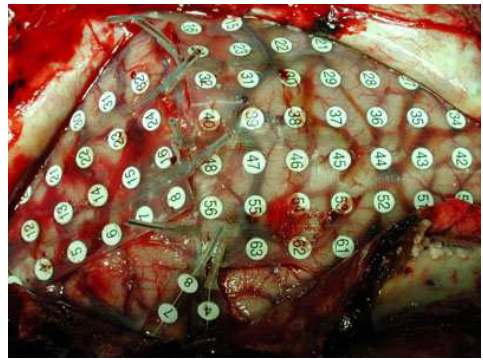}
\caption{(Color online)
Subdural ECoG grid of electrodes placed on
the surface of the brain for chronic evaluation of epileptic
patients before surgical resection.}\label{ichsa2}
\end{figure}

As part of the invasive surgical evaluation, arrays of electrodes are
chronically implanted onto the surface of the brain (Fig.\ref{ichsa2}) in order to record the onset and evolution of several
spontaneous seizures. The ensemble of signals [called the
electrocorticogram (ECoG)] is simultaneously recorded and analyzed for
detecting the proper cortical area to be resected \cite{schiff}.
Decisions regarding which areas of cortex to resect are
made by board-certified epileptologists, based on careful review of
the onset and spread of several spontaneous seizures.  The goal is to
remove the epileptogenic zone, the minimal area of cortex that is
necessary and sufficient to prevent further seizures from recurring,
if it is not located in eloquent cortical areas that are critical for
normal function.

The clinical challenge is how to analyze the ECoG signals to achieve a
simple, automatic, data-driven, fast and reliable identification of
the epileptic focus. Traditionally, linear measures, such as
correlation coefficients, were utilized to  complement the visual
inspection. In the last two decades methods based on the nonlinear
measures were introduced to the field.

 It has been shown that a significant reduction of the dimension of
the system \cite{CorDim,lehn01} (as measured by the correlation
dimension \cite{GP} and by the neuronal complexity loss) occur in the
vicinity of the epileptic focus, not only during the seizures
(ictal periods), but also between seizures (inter-ictal periods).

 The interdependence between signals from different sites, studied by
means of nonlinear cross-predictability and mutual nonlinear
predictability \cite{Arnold-InderDep,Queyen-interDep}, also allows the
location of the focus. Surrogate-corrected nonlinear measures
\cite{And06} were shown to correctly identify the focal hemisphere in
the inter-ictal recordings. Many of these methods involve the
reconstruction of the phase space by time-delay embedding
\cite{embedding}.  Such measures calculated from noisy
non-stationary time series usually can not be directly related to the
dimensionality of the attractor \cite{stam05} and the calculations are
computationally costly. 

Recently, concepts from synchronization analysis were brought to the
study of ECoG signals \cite{stam05,stam,quian,schindler,
schevon,bia06,palus01}, leading to a series of promising results on
localization and predictability of epileptic seizures. In most of
these works the increase in the synchronization was found only during
the seizure, while synchronization clusters, identified in
\cite{bia06}, remain synchronized also during inter-ictal
periods. Similar stability of local hyper-synchronous regions was
found also in \cite{schevon}. In spite of demonstrated success,
formulation of these methods involve some assumptions, that we would
like to relax in our proposed approach, as it will be explained in the
next Section.

The aim of this paper is twofold. First, we describe a model-free and
fast approach, based on phase synchronization, that is able to
identify the epileptic focus.  Second, we apply the method to ECoG
data of three patients, and show that there is a good correspondence
between the success or failure in surgery and the  overlap
between the area spanned by the electrodes identified by our method
and that of the brain tissue actually removed.
\section{Method}
\subsection{Patients and ECoG recordings}

 In this paper we analyze the ECoG data from three female patients
with intractable partial complex seizures with secondary
generalization, referred to as patients A, B and C, respectively.

 Patient A is a 17 years old and had her first seizure at age six.
Her magnetic resonance imaging (MRI) was normal, but a positron
emission tomography (PET) scan revealed reduced metabolism of the left
temporal and parietal lobes.  Spontaneous seizures originated in the
temporal lobe and spread posteriorly.  A total of 80 electrodes were
implanted in 4 arrays, with 16 eletrodes at the temporal lobe.  A left
temporal lobectomy was performed, leaving her seizure-free for the
last two years (Engel class I, see Ref. \cite{engel} for the
classification of the surgery outcome).

\begin{figure}
\includegraphics[width=0.45   \textwidth]{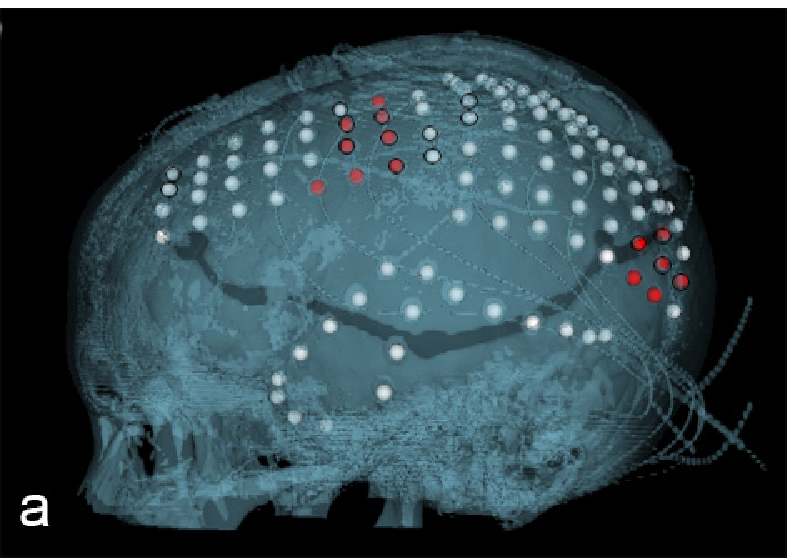}
\includegraphics[width=0.45   \textwidth]{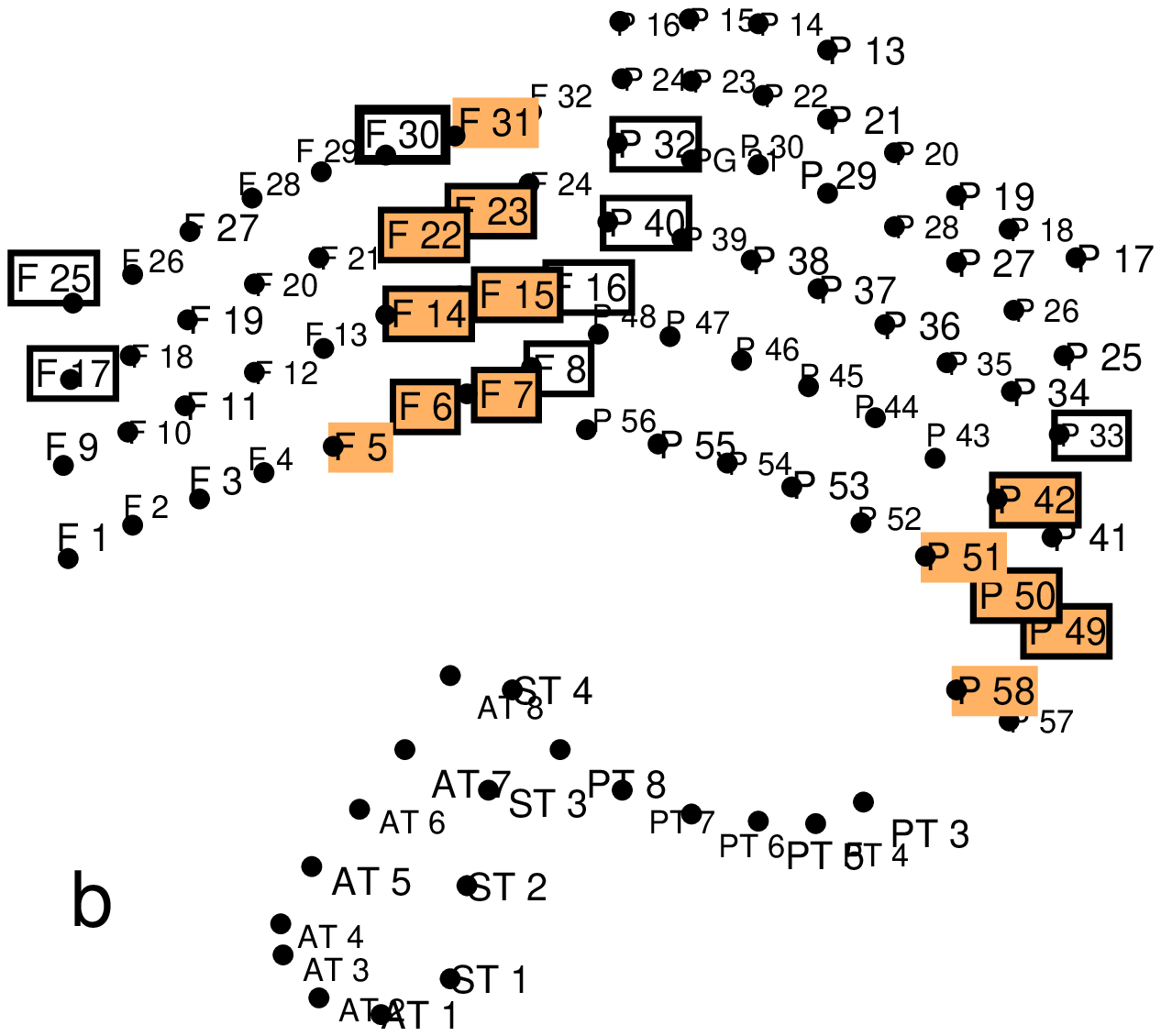}
\caption{(Color online) The computed tomography (CT) image with marked
electrodes (a) and the electrode scheme (b), projected on the plane,
for patient B.  The arrays of electrodes are labeled according to
their position: F- frontal, P- parietal, AT-anterior temporal, ST -
sub-temporal, PT - posterior-temporal. The same naming convention is
used for all patients. For description of other marks see discussion
after Eq. \ref{meanSbar}.}
\label{resb}
\end{figure}

 Patient B is a 12 years old with a history of developmental delay and
seizures since 8 month of age.  Before surgery she experienced as many
as 30 seizures/day.  Although her MRI was normal, a PET study revealed
general hypometabolism of the cerebral cortex.  96 electrodes in 5
arrays were implanted, with 32 electrodes in frontal grid (F), 32
electrodes in parietal grid (P) and 24 electrodes at the temporal lobe
in three strips (AT, ST and PT, cf. Fig. \ref{resb}). She received a
left frontal and occipital topectomy.  She had an Engel Class III
outcome.

 Patient C is a 7 years old and the outcome was Engel class II 16
months after operation, while 3 years after surgery she has frequent
seizures.   She was implanted with 96 electrodes, 64 of which were
in the parietal grid, and 32 electrodes at the temporal lobe.

 For all patients all subdural electrode channels were referenced to a
scalp electrode on the midline (CPz) and then the grand mean was
subtracted from each channel to remove the contribution of the
reference electrode.

We have analyzed a total of 24 hours of  ECoG recordings
(mean 8 hours, range 200 minutes-15 hours 30 minutes), containing a
total of 12 seizures (mean 4, range 3-5). The data were recorded with
a sampling frequency of 400 Hz and band-pass filtered offline to
0.5-50 Hz using moving overlapping windows and a finite impulse response
filter with a Kaiser window.

%%%%%%%%%%%%%%%%%%%%%%%
%%%%%%%%%%%%%%%%%%%%%%%%%%%%%

%%%%%%%%%%%%%%%%%%%%%%%%%%%%

%%%%%%%%%%%%%%%%%%%%%%%%%%%%%%%%%%%%%%%%%%%%%%%%%
  %%%%%%%%%%%%%%%%%%%%%%%%%%%%
\begin{figure}
\epsfig{width=.45\textwidth,file=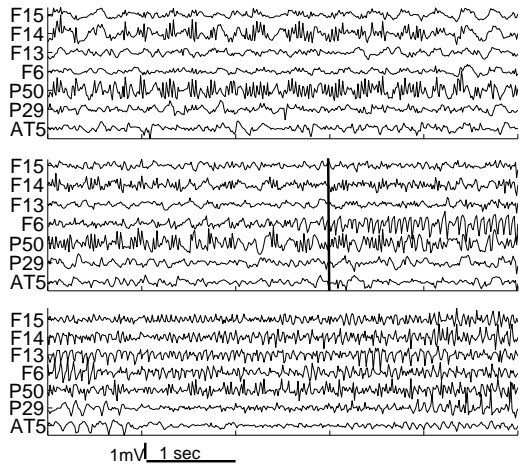}
 \caption{ Voltage traces of ECoG signals. Three 10 s portions of
voltage traces of 7 electrodes, taken from a multielectrode recording
of 96 electrodes of patient B,
cf. Fig. \ref{resb}. The seizure onset is marked by a vertical
line. The analysis below indicates that the signals coming from
electrodes F6, F14, F15 and P50 belong to the epileptic foci, whereas
the F13, P29 and AT5 do not. The voltage scale of 1 mV  and the time scale 
of 1 second are shown below the figure.}
 \label{signal}
 \end{figure}
 %%%%%%%%%%%%%%%%%%%%%%%%
\subsection{Synchronization Strength Diagrams}
To see the main concepts behind the method, consider a one-dimensional
oscillatory signal $x(t)$.  The associated phase space can be fully
reconstructed by using the coordinates $x(t)$, $dx(t)/dt$,
$d^2x(t)/dt^2,~\dots$, with as many derivatives as needed to exhaust
the dimensionality of the system. In such a reconstructed phase space
one can then define the Poincar\'e section by noting the points
obtained when the orbit crosses the surface $dx(t)/dt=0$. Obviously,
this surface coincides with the points of minima or maxima in the
original one-dimensional signal $x(t)$. In this work we focus on the
points of maxima, and define the {\em phase} $\phi(t)$ of the
trajectory by increasing its value by 2$\pi$ after each maximum.
Precisely,  between the $k$th maximum occurring at $t=t_k$ and the
$(k+1)$th maximum occurring at $t_{k+1}$, we define the phase over the period of
time $t_k \le t \le t_{k+1}$ using the linear interpolation
\cite{rose,bocca02}
\begin{equation}\label{defphase}
\phi(t) = 2\pi k + 2\pi \frac{t - t_{k}}{t_{k+1} - t_{k}}\quad (t_k <t<t_{k+1}) \ .
\end{equation}
Note that the resulting monotonically increasing phase does not
contain any information about the amplitude of the original signal and
thus the results of the  analysis are not affected by the amplitude
variation \cite{sharry}.

Phases can be affected by instrumental artifacts; we have carefully
ascertained that any signal segments that appear identical in all
recording channels were  manually excluded from the signal prior to calculation
of the phases.  It is important to remark that our data feature broad
band spectra. Alternative measures of the phase (such as the one
based on the Hilbert transform) imply the presence of a prominent
spectral peak \cite{bocca02}.
We would like to avoid any {\it a priori} assumptions on the spectral
structure of the signals, therefore we opted for the above definition
of phase, Eq. \ref{defphase}.

 %%%%%%%%%%%%%%%%%%%%%%%%%%%%
\begin{figure}
\includegraphics[width=0.45  \textwidth]{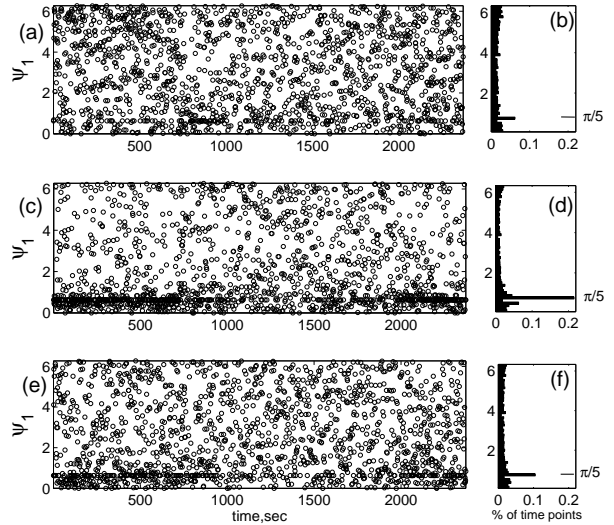}
 \caption{ Panels (a),(c) and (e): Synchrograms for 3 neighboring
pairs of electrodes along [F13-F14 (a), F15-F14(c)] and across the
grid [F6-F14 (e)], calculated for the reference electrode F14 (the
signals for F6, F13, F14 and F15 are shown in Fig.\ref{signal}). Since the
synchronization line appears at $\psi_1=0$, a small phase offset of
$\phi_{\rm sh}=\pi/5$ was added to the phase $\phi_y$ for
clarity. Panels (b),(d), and (f) show the histograms of the phase
points distribution within the corresponding synchrogram. Clearly,
only one strong line is present in each synchrogram, at the position
of the offset phase ($\pi/5$). This corresponds to the 1:1
synchronization without visible phase difference.}
 \label{synch1:1}
 \end{figure}
 %%%%%%%%%%%%%%%%%%%%%%%%

%%%%%%%%%%%%%%%%%%%%%%%%%%%%
\begin{figure}
\includegraphics[width=0.45  \textwidth]{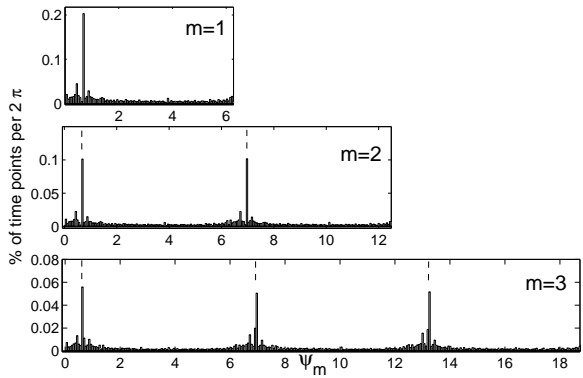}
 \caption{Histograms for synchrograms of orders $m=1,2$ and $3$
 calculated for the pair of electrodes F14-F15
 (cf. Fig. \ref{synch1:1}, c and d). The dashed lines mark offset
 positions $\pi/5, 2\pi+\pi/5$ and $4\pi+\pi/5$. The same bin size (
 $2\pi/100$) was used in all histograms. The number of strong lines
 equals to the order of the synchrogram, confirming the 1:1
 synchronization. The position of the synchronization lines coincide
 with the value of the offset phase within $\pm \pi/50$.}
 \label{synch1:2}
 \end{figure}
 %%%%%%%%%%%%%%%%%%%%%%%%
  %%%%%%%%%%%%%%%%%%%%%%%%%%%%
\begin{figure}

\includegraphics[width=0.45   \textwidth]{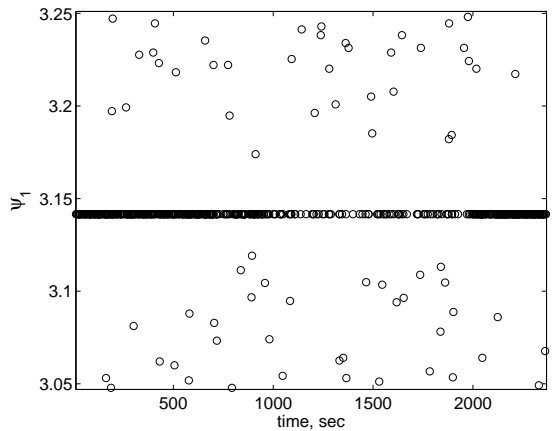}
 \caption{ The vicinity of the 1:1 synchronization line for
 synchrogram of electrodes F14 and F15, with the phase offset $\phi_{\rm sh}=\pi$. Note
 that the synchronization line is clearly isolated.}
 \label{synch1:3}
 \end{figure}
 %%%%%%%%%%%%%%%%%%%%%%%%

The notion of phase synchronization \cite{rose,bocca02} refers to a
process wherein a locking of the phases of weakly coupled
oscillators is produced, without implying a substantial
correlation in the oscillators' amplitudes. The same concept can
be related to our case, for measuring the degree of phase
synchronization of two different signals, say $x(t)$ and $y(t)$.
Using the basic definition (\ref{defphase}), one can indeed
produce the phases of each signal, denoted as $\phi_x(t)$ and
$\phi_y(t)$, respectively. Taking $x(t)$ as the reference signal,
one can further consider the time points of its maxima $t_k$, and
read the phase $\phi_y(t_k)$ at these time points. One then
defines the following reduced phases $\psi_m(t_k)$ according to
\begin{eqnarray}
\psi(t_k) &=&\phi_y(t_k) ~{\rm mod} 2 \pi \ , \nonumber \\
\psi_2(t_k)&=&\phi_y(t_k) ~{\rm mod} 4 \pi \ , \nonumber\\
&\dots &\ , \nonumber\\
\psi_m(t_k)&=&\phi_y(t_k) ~{\rm mod}~ 2\pi m \ . \label{reduced}
\end{eqnarray}
The graphic representation of $\psi_m(t_k)$ vs. $t_k$ is called a
synchrogram of order $m$. We will consider the signals $x(t)$ and
$y(t)$ as synchronized at ratio $n:m$ when the $m$th order synchrogram
exhibits $n$ distinct horizontal lines (i.e. fixed reduced phase
$\psi_m(t_k)$ for a series of time $t_k$).  Synchrograms, indeed,
have been proved to be useful to visualize and trace transitions
between different ratios of phase locking in biological data, as the
cardiorespiratory system of a human subject \cite{naturekurths}, where
non-stationarity is often strong.  Note that $\psi_m(t_k)$ is, in
general, not symmetric, and, within the same pair of signals, may
depend on which electrode is taken as a reference \cite{bocca02}.  An addition of
any constant phase $\phi_{\rm sh}$ to $\phi_y(t)$ does not change the
synchronization ratio, but merely shifts the position of the
synchronization lines. Such an offset may facilitate the visualization
of the line when it appears close to the edge of the synchrogram.

\begin{figure}
  \centering
\includegraphics[width=0.50   \textwidth]{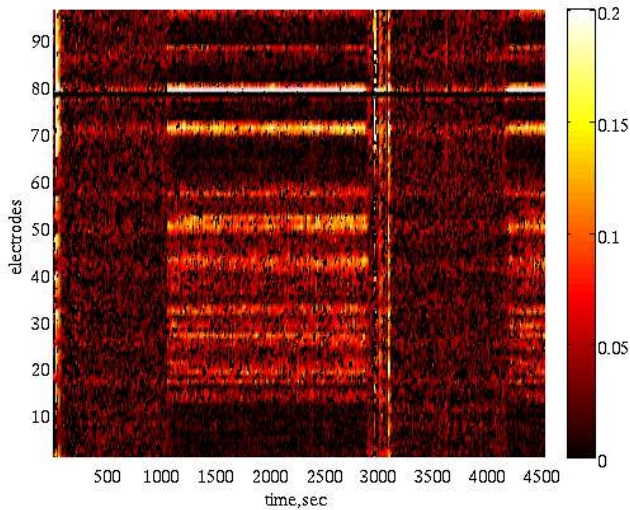}
\caption{(Color online) A typical SSD for a focal electrode [here, F14
(i=78)]. The electrodes here are numbered and correspond to: 1-8
(AT1-AT8), 9-12 (ST1-ST4), 13-58 (P13-P58), 59-64 (PT3-PT8), 65-96
(F1-F32). The time span of the diagram is 1 hour
30 minutes.}
\label{SSD1}
\end{figure}

\begin{figure}
  \centering
\includegraphics[width=0.50   \textwidth]{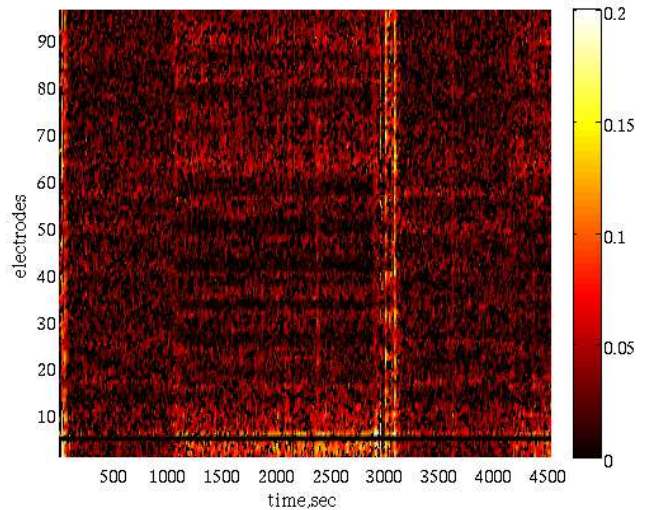}
\caption{(Color online) A typical SSD for a non-focal electrode [here,
AT5 (i=5)]. Electrodes are numbered as in Fig. \ref{SSD1}. The time span of the diagram is 1 hour
30 minutes. }
\label{SSD2}
\end{figure}

\begin{figure}
  \centering
\includegraphics[width=0.48   \textwidth]{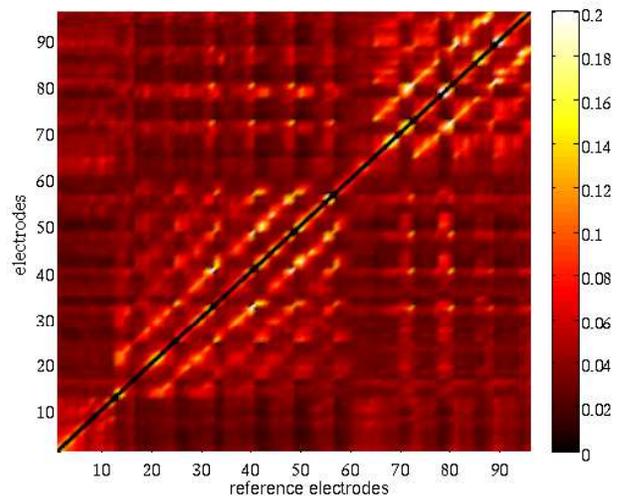}
\caption{(Color online) A typical mean pair-wise synchronization
matrix $\bar S_{i,j}$, computed from the same $S_{ij}$ as shown in
Figs. \ref{SSD1} and \ref{SSD2}, for the time interval (2000,2600)
sec.} \label{psm}
\end{figure}

To exemplify this notion in the present context, we use the data of
the patient B.  In Fig \ref{resb} we show the CT image with the
electrode positions marked on it and the plane projection of the
electrode placement scheme. In Fig. \ref{signal} the three 10 sec
portions of signals from seven electrodes are plotted. These portions
of the signal correspond to the time near the seizure onset. The
signals from five of these electrodes are used in
Figs. \ref{synch1:1}-\ref{psm}.

We present in Fig.~\ref{synch1:1}
synchrograms constructed from the signals associated with three
neighboring pairs of electrodes, which display the presence of 1:1
phase synchronization. The fraction of the time, during which the
signals are synchronized, differs significantly, albeit the same
geometrical distance  between these pairs of electrodes.

To test for the presence of more complicated synchronization ratios,
we calculated the synchrograms of higher orders. The histograms for
them are shown in Fig \ref{synch1:2}. The number of strong lines in
each histogram is equal to the order of the corresponding synchrogram,
thus confirming the dominant 1:1 synchronization ratio. Note that
this ratio is usually assumed in the methods involving phase
difference \cite{lehn01,schevon, bia06}. We do not make such an
assumption. In fact, any synchronization ratio may be treated on the
same footing.
%The analysis of long (several hours) recordings reveal
%occasional appearance of additional ratios of synchronization (1:2 and
%higher, always repeated as a group
%although they are weaker and do not affect the strength of the
%main 1:1 synchronization ratio.
 In the following analysis we
concentrate on the first order synchrograms $\psi_1$ with the phase offset
 $\phi_{\rm sh}=\pi$.  The phase points that belong to the synchronization line
may be easily extracted from the synchrogram, since they are clearly
isolated from other points (see Fig \ref{synch1:3}).

The present method shares some similarities with the so-called
event-synchronization technique \cite{quian}, though
this latter strategy is, in general, unable to distinguish between
different types of $n:m$ lockings.
%%%%%%%%%%%%%%%%%%%%%%%%%%%%%%
The information contained in the synchrograms can be easily employed to
extract the level of phase synchronization between all the pairs of
electrodes at our disposal. To this aim we define the  statistical strength of
synchronization $S_{ij}(t)$ (SS) of every given electrode $i$ to any other
electrode $j$: we take the synchrogram of electrodes $i$ and $j$, first
divide the time axis into windows (in our case of 10 seconds),
focusing on the line $\psi(t_k)=0$. In each such window, we calculate
the number $N_{ij}$ of phase points within the
range $\phi_{\rm sh} \pm 0.01$ . For $i\ne j$ $S_{ij}(t)$ is defined as $N_{ij}$
normalized by the total number of phase points throughout the measured
time interval, with $t$ being assigned to the first time-point in the
window. For $i=j$ we set $S_{ij}=0$ for all times. Note that this
procedure does not require time averaging and the choice of the time
window is dictated by the signal.  Precisely, if the time window
is too large, short events of strong synchronization will be smeared
out and not properly detected, thus reducing significantly the
sensitivity of the method. On the other hand, a too short time window
would result in a too small number of points over which the fraction
is calculated, thus introducing strong noise in the evaluation of the
synchronization strength. In the present case, we have chosen a window
length of 10 seconds, that represents a good compromise to obtain
maximum detail with minimal noise. The calculations were verified with
time windows of 1 and 20 seconds.

The calculation is repeated for all signals as a reference in
turn, since $S_{ij} \neq S_{ji}$ in general. The study of asymmetry is
beyond the scope of this paper.
%%%%%%%%%%%%%%%%%%%%%%%%%%%%
%%%%%%%%%%%%%%%%%%%%%%%%%%%%%%
%%%%%%%%%%%%%%%%%%%%%%%%%%%%
%%%%%%%%%%%%%%%%%%%%%%%%%%%%%%

%%%%%%%%%%%%%%%%%%%%%%%%%%%%
%%%%%%%%%%%%%%%%%%%%%%%%%%%%%%

  We refer to the diagram in which $S_{ij}(t)$ is displayed as the
Strength of Synchronization Diagram (SSD). We color code the values of
$S_{ij}(t) $ such that black  is zero and white  is the highest.

 Two such typical diagrams [using as reference electrodes the specific
electrodes F14 ($i=78$) and AT5 ($i=5$)] and $j$ from 1 to 96 are
presented in Figs. \ref{SSD1} and \ref{SSD2}. The diagrams
correspond to a part of the record one and a half hour long,
containing two subsequent seizures. The electrodes in these figures
are numbered and correspond to: 1-8 (AT1-AT8), 9-12 (ST1-ST4),13-58
(P13-P58), 59-64 (PT3-PT8), 65-96 (F1-F32) (cf. Fig \ref{resb}).
%  For the spatial relationship of these electrodes compare
%with Fig. \ref{resb} which shows the CT image and the two-dimensional
%projection of the actual placement of the electrodes in this case
%(patient B).

  Note that a ``strength of synchronization" of, say, 0.2,
means that the signals are synchronized 20\% of the time within the
given time window.

The SSD provides a clear scenario of synchronization dynamics between
these two seizures.
 The diagrams start with the final stages of the first
seizure (ending at about $t\approx 20$ seconds) with a high
synchronization strength in almost all channels. Then the
synchronization is lost for about 15 minutes. It then reappears at
$t\approx1100$ seconds.  In Fig. \ref{SSD1}, showing SSD for the
reference electrode F14, the most strong synchronization to the group
of neighboring electrodes F6-8 ($i=70-72$), F15-16 ($i=79-80$) and
F22-23 ($i=86-87$) is evident, while a larger group of electrodes is
synchronized with F14 at lower strength. Interestingly, this
electrode demonstrates similar level of synchronization strength with a
spatially remote P grid, especially with electrodes P49-PG52
($i=49-52$). At the same time, the electrode AT5 (Fig.\ref{SSD2})
becomes weakly synchronized with only few neighboring electrodes from
the same strip.  The next seizure occur between $t\approx 3000$
seconds and $t\approx 3240$ seconds.  Starting around $t=2940$ with a
{\em decrease} in synchronization the event proceeds to exhibit high
synchronization between all the electrodes, starting around
$t=3000$. After $t\approx 3240$ the synchronization is again lost and
regained by F14 at about $t=4300$ seconds, as it did at
$t\approx1100$.

 It is crucial to stress that the SSD patterns presented in
Figs. \ref{SSD1} and \ref{SSD2} are qualitatively preserved over all
seizures of the same patient in our data. Moreover, the groups of
strongly synchronized electrodes are preserved throughout all records,
%\underline {both when the patient B was awake and asleep,}
 although at varying strength. The desynchronization following the
seizure, apparent in Figs. \ref{SSD1} and \ref{SSD2}, corresponds to
the post-ictal state, detected at the time of recording.

Similar behavior was found in the data of two other patients.  In a
patient A's data only four pairs of electrodes remain synchronized
during all available records.  Two of these pairs were placed at the
temporal lobe. During the seizures, a particularly strong
synchronization was found in the signals coming only from AT and ST electrodes.
For patient C three independent groups of strongly synchronized
electrodes were found, while the overall behavior was similar to that
of patient B.

\section{Results and Discussion}

%%%%%%%%%%%%%%%%%%%%%%%%%%%%
\begin{figure}
 \includegraphics[width=0.45   \textwidth]{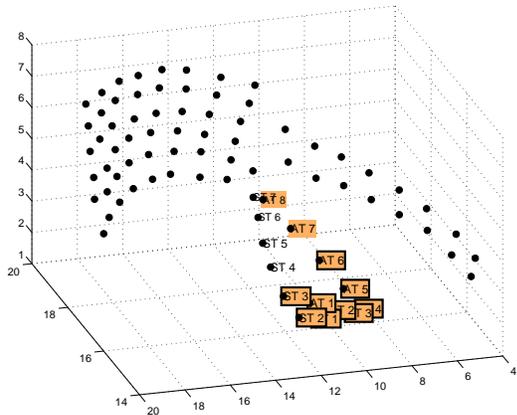}
\caption{(Color online) The geometric spread of the electrodes, for
the patient A. The viewing angle was chosen to best present the
resected area. Only relevant electrodes are labeled. The electrodes
with synchronization strength exceeding the threshold value $\bar S$
are marked by a rectangle. The electrodes chosen for resection are
highlighted.  }
\label{resa}
\end{figure}

%%%%%%%%%%%%%%%%%%%%%%%%%%%%%%%%%%%%%%%%%%%%%%%%%
%%%%%%%%%%%%%%%%%%%%%%%%%%%%%%%%%%%%%%%%%%%%%%%%%

\begin{figure}

 \includegraphics[width=0.45   \textwidth]{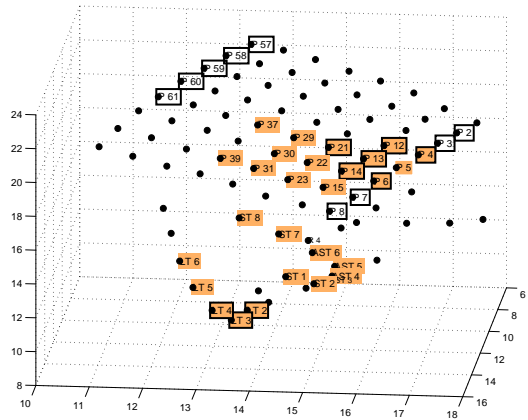}
\caption{(Color online) The electrodes scheme for the patient C. The
viewing angle was chosen to best present the resected areas. Only
relevant electrodes are labeled for clarity. The electrodes are marked
as in Fig \ref{resa}. }
\label{resc}
\end{figure}

%%%%%%%%%%%%%%%%%%%%%%%%%%%%%%%%%%%%%%%%%%%%%%%%%
 Previous observations involving different linear and non-linear measures
\cite{bia06, schevon, towle99, doron06, quyen05} indicate that the
values of the corresponding measures are altered over tumors and
regions thought to be part of the epileptogenic zone. The regions of
such a locally enhanced or suppressed activity are stable and have
well defined edges.  Since we found high synchronization strength
during epileptic seizures, we identify the foci of epileptic activity
with those regions that  statistically preserve a high degree of
synchronization at all times. Thus our definition of the foci of 
epileptic seizures is the ensemble of those electrodes that, besides
being strongly synchronized to each other during the ictal period,
remain also synchronized above a chosen threshold during the
inter-ictal period. To identify this ensemble, we now compute the time
average of $ \bar S_{ij}$ according to
\begin{equation}
\bar S_{ij} \equiv T^{-1} \sum_{t_k=1}^{T} S_{ij}(t_k) \ ,\label{barS}
\end{equation}
where the time interval $(1,T)$ does not include a seizure. An example
of the resulting mean pair-wise synchronization matrix $\bar S_{ij}$
is shown in Fig. \ref{psm}. $\bar S_{i,j}$ converges well for $T>60$
seconds.  Clearly, the synchronization strength between electrodes within a given grid
is higher than between spatially remote electrodes.   For a collection
${\cal E}_{\rm ref}$ of $N_{\rm ref}$ reference electrodes and a
collection ${\cal E}_{\rm el}$ of $N_{\rm el}$ electrodes
(cf. Fig. \ref{psm}) we define the mean $\bar S$ as:
\begin{equation}
\bar S \equiv \frac{1}{N_{\rm el} N_{\rm ref} }\sum_{i \in {\cal E}_{\rm ref}}\sum_{j \in {\cal E}_{\rm el}} \bar S_{ij} \ .\label{meanSbar}
\end{equation}
 For the diagram shown in Fig. \ref{psm}, we calculated $\bar S$
and the standard deviation $\sigma$ of $\bar S_{ij}$ for each of five
grids and between the grids.  The value of $\bar S$ within the grid,
averaged over five grids is $0.056$ (varying $0.047-0.065$ for
different grids). The standard deviation $\sigma$ within the grid,
averaged over five grids is $\sigma=0.034$ (ranging $0.03-0.042$ for
different grids).  The value of $\bar S$ averaged over all inter-grid
combination of electrodes is $0.038$ (range $0.031-0.045$) within
$\sigma=0.007$ (range $0.004-0.0125$).
% The mean value of $\bar S$ within the grid is $0.056$ (range
%$0.047-0.065$) within $\sigma=0.01$ (range $0.0058-0.013$), while the
%mean inter-grid $\bar S$ is $0.038$ (range $0.031-0.04$) within
%$\sigma=0.004$ (range $0.002-0.006$)} .
At the same time, the SS of the strongly synchronized group ranges in
the interval $0.1-0.2$, being more than three standard deviations
above the corresponding mean value. The difference is especially
striking for the inter-grid synchronization, found between the group
in F grid (F6-7, F14-16), and the remote group in P grid (P49, P50).
 These values are typical for all data of patient B, except for
post-ictal periods. Similar values of $\bar S_{ij}$ were found for
patients A and C, although no such remote synchronization was
detected.  This observation is consistent with the relations between values of
mean phase coherence of the electrodes pairs that belong to the local
hypersynchrony regions and those that are outside of these regions, found in \cite{schevon}.

To localize the foci we need to choose a threshold value $\bar S_{\rm
th}$, such that only electrode pairs with $\bar S_{ij}> \bar S_{\rm
th}$ will be selected as belonging to the focus.  Although for
these pairs usually both $\bar S_{ij}$ and $\bar S_{ji}$ were above
the threshold, we always used the largest of the two to compare with
$\bar S_{\rm th}$.  Clearly, the number of selected electrodes is a
strong function of $\bar S_{\rm th}$. To compare with the actual
selection we chose $\bar S_{\rm th}$, such that the number of
electrodes designated by the present procedure agrees as much as
possible with the number chosen by the epileptologists.  It turned out
that chosen values of $\bar S_{\rm th}$ are larger than the value of
$\bar S$ by at least three standard deviations.

The results, that refer to calculations over the entire set of
available data, are displayed in Figs. \ref{resb}, \ref{resa} and \ref{resc}.  We color
those electrodes that were chosen  for the operation,
and we put a rectangle on every electrode that is identified by our
method.

For patient A, the SS was relatively low in the inter-ictal periods.
%with only two pairs of electrodes strongly synchronized.
 On the other hand, during the seizure, the entire region, covered by
AT and ST strips was synchronized at a level of 0.6-0.7. We included
in the selection 9 electrodes, that showed highest  pairwise
synchronization. They all are placed within the resected area, see
Fig. \ref{resa}. This patient was seizure-free after the surgery.

We have shown in Figs \ref{SSD1}-\ref{psm} that patient B had two
remote groups of electrodes that remained inter-ictally
synchronized. As seen in Fig \ref{resb}, the chosen subset of strongly
synchronized electrodes overlap with both foci, resected during
surgery. Out of 13 electrodes marked for resection, 9 were identified
also by us. However, we have found also other electrodes, that belong
to the same focus (in the frontal grid), as well as adjacent electrodes
from the parietal grid. The outcome of the surgery was a improvement
in patient's condition, who however did not become seizure-free.

As for patient C (cf. Fig.\ref{resc}), the operated area and our
chosen set of electrodes overlap only partially. Out of 30 electrodes
marked by epileptologists, only 9 were chosen also by us. An
additional strongly synchronized group was found at the edge of the parietal
grid, indicating the possibility for another, not resected
focus. After initial improvement, this patient did not benefit from
the surgery.

In summary, we pointed out that a simple and computationally fast
method to detect phase synchrony in ECoG signals was able to detect
and localize the multiple foci of epilepsy. The method does not
require any assumptions about the spectral features of the signal or
the nature of the synchronization between the signals. The measurement
of the synchronization strength from the synchrogram amounts to a
straightforward count of the data points within a chosen
interval.  The synchronization scenario in the
long recordings is characterized by repeated patterns from one seizure
to the next one.  For each patient, we found a subset of electrodes,
that statistically preserve a high degree of synchronization at
all times. These subsets overlap fully or partially with the areas
chosen for resection. In patient B we correctly identified the
existence of two distant foci. In patient C we also identified the
presence of additional focus, not operated on.  The overlap of the
location of the electrodes identified by our method and the resected
areas was consistent with the degree of surgery success.

Although it is not reasonable to draw any final conclusions on the basis
of the study of only three patients, these observations suggest that this method
 may become a useful tool to help improve the outcome of the surgical treatment of epilepsy.

\section*{Aknowledgments}
We thank Itai Doron and Tomer Gazit for fruitful
discussions.  This work was partially supported by a EU grant under
the GABA project.  S.B. acknowledges a visiting fellowship granted by
the Yeshaya Horowitz Association through the Center for Complexity
Science.

\end{document}